\documentclass[a4paper,12pt]{article}
\usepackage{amsmath,amsfonts,amssymb}
\usepackage{graphicx}
\usepackage[export]{adjustbox}
\usepackage{multirow}
\usepackage[noadjust]{cite}
\usepackage{scrextend}
\usepackage[singlelinecheck=false]{caption}

\newcommand{\gmv}{\langle\left(G_{\mu\nu}^a\right) ^2\rangle}
\newcommand{\qqb}{\langle\bar{q}q\rangle}
\newcommand{\als}{\alpha_s}
\newcommand{\lbbr}{\left(}
\newcommand{\rbbr}{\right)}
\newcommand{\wtm}{\widetilde{m}^2(n)}
\newcommand{\bc}{\bar{C}}
\newcommand{\bmb}{\bar{M}}

\begin{document}

\begin{center}
{\Large\bf Large-$N_c$ masses of light mesons from QCD sum rules for non-linear radial Regge trajectories}
\end{center}
\bigskip
\begin{center}
{\large S. S. Afonin
and T. D. Solomko}
\end{center}

\begin{center}
{\it Saint Petersburg State University, 7/9 Universitetskaya nab.,
St.Petersburg, 199034, Russia}
\end{center}
\bigskip

\begin{abstract}
The large-$N_c$ masses of light vector, axial, scalar and pseudoscalar mesons
are calculated from QCD spectral sum rules for a particular ansatz
interpolating the radial Regge trajectories. The ansatz includes a linear part
plus exponentially degreasing corrections to the meson masses and residues.
The form of corrections was proposed some time ago from consistency with
analytical structure of Operator Product Expansion of the two-point correlation
functions. Two solutions are found and compared with the experimental data.
\end{abstract}

\section{Introduction}

The hadron spectrum is crucially shaped by the confinement property of QCD.
It is often believed that in the sector of light mesons this property
should lead to emergence of approximately linear Regge and radial trajectories
with nearly universal slope. The hadron phenomenology seems to agree with
this global feature of light meson spectrum,
at least qualitatively~\cite{phen,phen2,phen3,phen4,phen5,phen6,phen7,phen8}.
The slope of angular and radial trajectories becomes a highly
important quantity appearing from the non-perturbative QCD and giving rise to the
scale of hadron masses in the light quark sector. Another important quantity
in this picture represents the intercept which strongly depends on quantum
numbers of a particular trajectory. The real meson spectrum reveals also
deviations from the linear trajectories, sometimes quite noticeable.
A pattern of these non-linear corrections remains obscure.

The problem of deciphering the general structure of light meson spectrum
can be addressed by different methods. A fruitful approach closely related with QCD
is the method of planar QCD sum rules (see, e.g., a short review in Ref.~\cite{AS}).
It is based on merging the ideas of classical SVZ sum rules~\cite{svz} and
the large-$N_c$ (often called planar) limit in QCD~\cite{hoof,witten}.
Within the given method, the problem of description of non-linear corrections
to the straight radial trajectories was studied in detail in Ref.~\cite{we}.
In view of many phenomenological developments in the hadron spectroscopy
in the last fourteen years we have found useful to check critically the
conclusions made in Ref.~\cite{we} and try to use different assumptions
in the proposed model. This constitutes the main purpose of the present work.

The paper is organized as follows. In Section~2, we formulate our model and
derive the ensuing planar sum rules in the vector and scalar channels. Some
insignificant errors noticed in Ref.~\cite{we} are corrected.
The obtained equations are numerically solved in Section~3. The main new result here
is finding the second solution which was missed in the original paper~\cite{we}.
Our attempts to use alternative assumptions are briefly described in Section~4.
We conclude in Section~5.

\section{Sum rules}
\subsection{The two-point correlators}

The QCD sum rules stem from the Operator Product Expansion (OPE) of two-point correlators of
various quark currents in Euclidean space~\cite{svz,rry} (see also the review~\cite{colangelo}),
\begin{equation}
\label{1}
\Pi^J(Q^2)=\int d^4x\,e^{iQx}\langle\bar{q}\Gamma q(x)\bar{q}\Gamma q(0)\rangle,
\end{equation}
where $Q$ denotes the Euclidean momentum and we will consider the scalar (S), pseudoscalar (P),
vector (V) and axial-vector (A) channels, i.e. $J=S,P,V,A$ corresponding to
$\Gamma=i,\gamma_5,\gamma_\mu,\gamma_\mu \gamma_5$.
The scalar part of vector and axial correlators is defined by
\begin{equation}
\label{2}
\Pi_{\mu\nu}^{V,A}=\lbbr -\delta_{\mu\nu} Q^2 +Q_\mu Q_\nu \rbbr \Pi^{V,A}(Q^2).
\end{equation}
The OPE for these correlators at one-loop level and in the chiral limit reads~\cite{svz,rry}
\begin{equation}
\label{3}
\Pi^V(Q^2)=\frac{1}{4\pi^2}\lbbr 1+\frac{\als}{\pi}\rbbr \ln\frac{\mu^2}{Q^2}+\frac{\als}{12 \pi}\frac{\gmv}{Q^4} -\frac{28}{9}\pi\als\frac{\qqb^2}{Q^6},
\end{equation}
\begin{equation}
\label{4}
\Pi^A(Q^2)=\frac{1}{4\pi^2}\lbbr 1+\frac{\als}{\pi}\rbbr \ln\frac{\mu^2}{Q^2}+\frac{\als}{12 \pi}\frac{\gmv}{Q^4} +\frac{44}{9}\pi\als\frac{\qqb^2}{Q^6},
\end{equation}
\begin{equation}
\label{5}
\Pi^S(Q^2)=-\frac{3}{8\pi^2}\lbbr 1+\frac{11\als}{3\pi}\rbbr Q^2\ln\frac{\mu^2}{Q^2}+\frac{\als}{8 \pi}\frac{\gmv}{Q^2} -\frac{22}{3}\pi\als\frac{\qqb^2}{Q^4},
\end{equation}
\begin{equation}
\label{6}
\Pi^P(Q^2)=-\frac{3}{8\pi^2}\lbbr 1+\frac{11\als}{3\pi}\rbbr Q^2\ln\frac{\mu^2}{Q^2}+\frac{\als}{8 \pi}\frac{\gmv}{Q^2} +\frac{14}{3}\pi\als\frac{\qqb^2}{Q^4},
\end{equation}
where $\qqb$ and $\gmv$ mean the quark and gluon condensate respectively
and the further inverse in $Q^2$ terms are omitted. The numerical coefficients
at the condensate contributions are given in the large-$N_c$ limit of QCD~\cite{hoof,witten}.

On the other hand, in the planar limit, the two-point correlators have the following resonance representation,
\begin{equation}
\label{7}
\Pi^J(Q^2)=\sum_{n=0}^{\infty}\frac{Z_J(n)}{Q^2+m_J^2(n)},
\end{equation}
where $n$ denotes the number of radial excitation.

The usual planar QCD sum rules are obtained after summing in $n$, decomposing the result in $Q^{-2}$,
and matching to the corresponding OPE.

\subsection{The vector and axial cases}

Following the motivation of Ref.~\cite{we} (see also~\cite{nph})
we consider the following form for the non-linear radial spectrum,
\begin{align}
\label{8}
m^2_{V,A}(n)&=M^2+a n+A_m^{V,A} e^{-B_m n}, \\
F^2_{V,A}(n)&=a\lbbr C+A_F^{V,A} e^{-B_F n}\rbbr.
\label{9}
\end{align}
Here $F^2_{V,A}(n)\equiv Z_{V,A}(n)/2$.
The given ansatz consists from the linear part plus an exponentially decreasing correction
(the physical spectrum corresponds to $B_m > 0$, $B_F > 0$).
The form of this correction was dictated by the requirement to reproduce the
analytical structure of OPE in the variable $Q^2$~\cite{we}.

To avoid the irrelevant infinite constants we will consider the first derivative in $Q^2$.
Introducing the notation
\begin{equation}
\label{10}
\bar{m}^2(n)=M^2+an,
\end{equation}
we obtain
\begin{equation}
\label{11}
\frac{d\Pi(Q^2)}{dQ^2}=-2\sum_{n=0}^\infty\frac{a\lbbr C+A_F e^{-B_F n}\rbbr }{\lbbr Q^2+\bar{m}^2(n)+A_m e^{-B_m n}\rbbr ^2}.
\end{equation}
This expression is not analytically summable and we need some approximation.
Since the exponential contribution is presumably small except the ground state $n=0$,
the further strategy is to keep only linear in exponential contribution terms for the
excited states while the ground states are taken into account exactly. Below we
display the resulting sum rules after matching to the OPE~\cite{we}.

\noindent The sum rule at $1/Q^2$,
\begin{equation}
\label{12}
\frac{1}{8 \pi^2}\lbbr 1+\frac{\alpha_s}{\pi}\rbbr =C.
\end{equation}
The sum rules at $1/Q^4$,
\begin{align}
a\lbbr C+A_F^V\rbbr -C\lbbr \frac{a}{2}+M^2\rbbr +A_F^V\Delta_F^{(1)}&=0,\\
a\lbbr C+A_F^A\rbbr -C\lbbr \frac{a}{2}+M^2\rbbr +A_F^A\Delta_F^{(1)}&=-f_\pi^2.
\end{align}
The sum rules at $1/Q^6$,
\begin{multline}
-2a\lbbr C+A_F^V\rbbr \lbbr M^2+A_m^V\rbbr +C\lbbr M^4+M^2 a+\frac{a^2}{6}\rbbr\\
-2 C A_m^V \Delta_m^{(1)}-2 A_F^V \Delta_F^{(2)}=\frac{\alpha_s}{12\pi}\gmv,
\end{multline}
\begin{multline}
-2a\lbbr C+A_F^A\rbbr \lbbr M^2+A_m^A\rbbr +C\lbbr M^4+M^2 a+\frac{a^2}{6}\rbbr\\
-2 C A_m^A \Delta_m^{(1)}-2 A_F^A \Delta_F^{(2)}=\frac{\alpha_s}{12\pi}\gmv.
\end{multline}
The sum rule at $1/Q^8$ in $\Pi^V(Q^2)-\Pi^A(Q^2)$,
\begin{multline}
3a\lbbr C+A_F^V\rbbr \lbbr M^2+A_m^V\rbbr ^2-3a\lbbr C+A_F^A\rbbr \lbbr M^2+A_m^A\rbbr ^2\\
+6 C \lbbr A_m^V-A_m^A\rbbr \Delta_m^{(2)}+3(A_F^V-A_F^A)\Delta_F^{(3)}=-12\pi \alpha_s \qqb^2.
\label{17}
\end{multline}
We introduced the following notations in the relations above,
\begin{align}
\Delta_i^{(1)}&=\frac{a}{e^{B_i}-1},\\
\Delta_i^{(2)}&=\frac{a\lbbr -M^2+\lbbr M^2+a\rbbr e^{B_i}\rbbr }{\lbbr e^{B_i}-1\rbbr ^2},\\
\Delta_i^{(3)}&=\frac{a\lbbr -a\lbbr a+2 M^2\rbbr +a e^{B_i}\lbbr 2M^2+3a\rbbr +\lbbr M^2+a\rbbr ^2\lbbr e^{B_i}-1\rbbr ^2\rbbr }{\lbbr e^{B_i}-1\rbbr ^3}.
\end{align}

\subsection{The scalar and pseudoscalar cases}

We define the scalar residues as
\begin{equation}
\label{21}
Z_{S,P}(n)=2G^2_{S,P}(n)m^2_{S,P}(n).
\end{equation}
For the $\pi$-meson, this definition makes sense if it belongs to the pseudoscalar radial trajectory.
It looks more likely, however, that it does not belong to this trajectory because of its
pseudogoldstone nature. In this case, we should use a current algebra relation $Z_\pi=2\frac{\qqb^2}{f_{\pi}^2}$~\cite{we}.
We will consider the both cases and refer to them as $\pi$-in è $\pi$-out correspondingly.
Similarly to the vector channels, the scalar spectrum will be interpolated by the following ansatz~\cite{we},
\begin{align}
\label{22}
m^2_{S,P}(n)&=\bmb ^2+a n+A_m^{S,P} e^{-B_m n}, \\
G^2_{S,P}(n)&=a\lbbr \bc +A_G^{S,P} e^{-B_G n}\rbbr.
\label{23}
\end{align}
As was motivated in Ref.~\cite{we} the slope $a$ and exponent $B_m$ are considered as universal parameters
determined mainly by pure gluodynamics.

Let us introduce the notation
\begin{equation}
\label{24}
\wtm=\bmb ^2+an,
\end{equation}
and consider the cases $\pi$-out and $\pi$-in separately.
To avoid the infinite contact terms, the second derivatives of the scalar correlators, $\frac{d^2\Pi(Q^2)}{(dQ^2)^2}$, will be analyzed.
Using the procedure described above for the vector mesons we obtain the following set of sum rules for the case $\pi$-out~\cite{we}.

\noindent The sum rule at $1/Q^2$,
\begin{equation}
\label{25}
\frac{3}{32 \pi^2} \lbbr 1+\frac{11 \alpha_s}{3 \pi}\rbbr =\frac{\bc }{2}.
\end{equation}
The sum rules at $1/Q^6$,
\begin{multline}
a(\bc +A_G^S)(\bmb ^2+A_m^S)-\frac{\bc }{2}\lbbr \bmb ^2(\bmb ^2+a)+\frac{a^2}{6}\rbbr \\ + A_G^S \Delta_G^{(2)}(\bmb )+ \mathbf{\frac{a \bc  A_m^S}{e^{B_m}-1}} = \frac{\alpha_s}{16 \pi} \gmv,
\end{multline}
\begin{multline}
\frac{\qqb^2}{f_\pi^2}+a(\bc +A_G^P)(\bmb ^2+A_m^P)-\frac{\bc }{2}\lbbr \bmb ^2(\bmb ^2+a)+\frac{a^2}{6}\rbbr \\ + A_G^P \Delta_G^{(2)}(\bmb )+ \mathbf{\frac{a \bc  A_m^P}{e^{B_m}-1}} = \frac{\alpha_s}{16 \pi} \gmv.
\end{multline}
The sum rule at $1/Q^8$ in $\Pi^S(Q^2)-\Pi^P(Q^2)$,
\begin{multline}
\label{28}
-3a(\bc +A_G^S)(\bmb ^2+A_m^S)^2+3a(\bc +A_G^P)(\bmb ^2+A_m^P)^2\\-\mathbf{6}\bc (A_m^S-A_m^P)\Delta_m^{(2)}(\bmb )-3(A_G^S-A_G^P)\Delta_G^{(3)}(\bmb )=-18 \pi \alpha_s \qqb^2.
\end{multline}

In these sum rules and in relations below, we write in bold face the contributions which were missed in Ref.~\cite{we}.

The sum rules for the $\pi$-in case are the same but the terms containing the factor $(\bc +A_G^P)$ are absent.

\subsection{Chiral constants and electromagnetic pion mass difference}

If the spectral parameters are obtained from the sum rules we can calculate
the important constants $L_{10}$ and $L_{8}$ of $SU(3)$ chiral perturbation theory~\cite{gasser}
(the corresponding phenomenological values are $L_{10}=(-5.5\pm0.7)\cdot10^{-3}$ and $L_8=(0.8\pm0.3)\cdot10^{-3}$)
and the electromagnetic pion mass difference $\Delta m_\pi \equiv m_{\pi^+}-m_{\pi^0}$ (see, e.g., Ref.~\cite{pan}
for a short review), its experimental value is $\Delta m_\pi=4.59\,\text{MeV}$~\cite{pdg}.

The corresponding chiral constants are defined by the relations
\begin{align}
L_{10}&=-\frac{1}{8}\frac{d}{dQ^2}\lbbr Q^2\lbbr \Pi^V(Q^2)-\Pi^A(Q^2)\rbbr \rbbr \Bigr|_{\substack{Q^2=0}},\\
L_8&=\frac{f_{\pi}^4}{32 \langle\bar{q}q\rangle^2}\frac{d}{dQ^2}\lbbr Q^2\lbbr \Pi^S(Q^2)-\Pi^P(Q^2)\rbbr \rbbr \Bigr|_{\substack{Q^2=0}}.
\end{align}
Substituting our ansatz and using our approximations we obtain
\begin{multline}
L_{10}=\frac{a}{4}\lbbr \frac{C+A_F^A}{M^2+A_m^A}-\frac{C+A_F^V}{M^2+A_m^V}\right. \\
\left. +\sum_{n=1}^\infty \frac{\bar{m}^2(n) e^{-B_F n} \lbbr A_F^A-A_F^V\rbbr -C e^{-B_m n} \lbbr A_m^A-A_m^V\rbbr }{\bar{m}^4(n)}\rbbr,
\end{multline}
\begin{align}
L_8&=\frac{f_{\pi}^4}{16 \langle\bar{q}q\rangle^2}\lbbr \bc+A_G^S+\frac{A_G^S-A_G^P}{e^{B_G}-1}\rbbr ,\quad\text{case }\pi\text{-in},\\
L_8&=\frac{f_{\pi}^4}{16 \langle\bar{q}q\rangle^2}\frac{A_G^S-A_G^P}{1-e^{-B_G}},\quad\text{case }\pi\text{-out}.
\end{align}

The electromagnetic pion mass difference is given by
\begin{equation}
\label{37}
\Delta m_\pi = \frac{3 \alpha}{16 \pi m_\pi f_\pi^2}\int_0^\infty dQ^2 Q^2 \lbbr\Pi^A(Q^2)-\Pi^V(Q^2)\rbbr.
\end{equation}
Here $\alpha$ denotes the fine structure constant.
This formula leads to the following expression~\cite{we},
\begin{multline}
    \Delta m_\pi = \frac{3 \alpha}{8 \pi m_\pi f_\pi^2} \left\{\vphantom{\frac{m_A^2(n)}{m_V^2(n)}}(C+A_F^A)(M^2+A_m^A)\ln\lbbr M^2+A_m^A\rbbr\right.\\
    -(C+A_F^V)(M^2+A_m^V)\ln\lbbr M^2+A_m^V\rbbr + \sum_{n=1}^\infty \left[ C \bar{m}^2(n)\ln\frac{m_A^2(n)}{m_V^2(n)}\right. \\
     +\left.\left.\lbbr C e^{-B_m n} (A_m^A-A_m^V)+\bar{m}^2 (n) e^{-B_F n} (A_F^A-A_F^V) \rbbr \ln \bar{m}^2(n)\right]\right\},
\end{multline}
where the dimensional quantities under the logarithms must be divided by arbitrary scale $\mu^2$. The result does not depend on $\mu^2$
due to the imposed sum rules. The calculation of Ref.~\cite{we}, however, was not carried out to its logical end: We keep only the
linear in exponential corrections terms and this was not done for the first logarithm in the infinite sum. The correct calculation results
in the final expression
\begin{multline}
    \Delta m_\pi = \frac{3 \alpha}{8 \pi m_\pi f_\pi^2} \left\{\vphantom{\frac{m_A^2(n)}{m_V^2(n)}}(C+A_F^A)(M^2+A_m^A)\ln\lbbr M^2+A_m^A\rbbr \right. \\
    -(C+A_F^V)(M^2+A_m^V)\ln\lbbr M^2+A_m^V\rbbr + \frac{C (A_m^A-A_m^V)}{e^{B_m}-1} \\
    \left. +\sum_{n=1}^\infty \lbbr C e^{-B_m n} (A_m^A-A_m^V)+\bar{m}^2 (n) e^{-B_F n} (A_F^A-A_F^V) \rbbr \ln \bar{m}^2(n)\right\}.
\end{multline}

\section{Numerical results}

Having a set of sum rules and fixing some inputs one can calculate the mass spectrum in each channel.
As inputs we take the masses of ground states $m_V(0)$, $m_A(0)$, $m_S(0)$, $m_P(0)$, and of first pseudoscalar excitation $m_P(1)$.
Other inputs are taken as in Ref.~\cite{we}: $a=(1120\,\mbox{MeV})^2$, $\langle\bar
qq\rangle=-(240\,\mbox{MeV})^3$,
$\frac{\alpha_s}{\pi}\langle\left(G_{\mu\nu}^a\right)^2\rangle=
(360\,\mbox{MeV})^4$, $f_{\pi}=103\,\mbox{MeV}$,
$Z_{\pi}=2\frac{\langle\bar qq\rangle^2}{f_{\pi}^2}$,
$\alpha_s=0.3$. The units are: $m(n)$, $F(n)$, $G(n)$ --- MeV; $A_{m}$
--- MeV${}^2$; $A_F$, $A_G$, $B_{F,G,m}$ --- MeV${}^{0}$.

In comparing with phenomenological values we use the data from PDG~\cite{pdg}
and try to neglect states with large admixture of $s$-quark and $D$-wave vector mesons.

\subsection{Vector and axial-vector mesons}

The mass spectrum ansatz~\eqref{8}--\eqref{9} contains 9 parameters. We have 6 sum rules~\eqref{12}--\eqref{17}.
The slope $a$ is fixed from the phenomenology. Thus we need 2 additional constraints.
These constraints will be the masses of ground states whose values are taken as in Ref.~\cite{we}:
$m_V(0)=770\,\text{MeV}$, $m_A(0)=1200\,\text{MeV}$.

We found two numerical solutions for the system of equations~\eqref{12}--\eqref{17} supplemented by 2 additional constraints,
they are displayed in Tables~1-4 and Figs.~1,2. The first one is close to the solution found in Ref.~\cite{we}
(where, e.g., $B_m=0.97$ against our $B_m=1.024$). A small difference is just due to a better precision
of the present numerical calculation. To our surprise, there exists the second solution which was completely
missed in Ref.~\cite{we}. The second solution corresponds to $B_m=0.392$. It is seen that
the experimental data are better described by the second solution.

It is worth noting that the electromagnetic pion mass difference $\Delta m_\pi$ is very sensitive
to the values of inputs $m_V(0)$ and $m_A(0)$. In fact, our choice above, which slightly differs from the
corresponding central experimental values $m_V(0)=775\,\text{MeV}$ and $m_A(0)=1230\,\text{MeV}$~\cite{pdg},  was dictated
by our wish to reproduce $\Delta m_\pi$ close to its experimental value (with those central values $\Delta m_\pi$ would be equal to
$18\,\text{MeV}$ in the first solution and $39\,\text{MeV}$ in the second one). In this sense, $\Delta m_\pi$ represents
rather an input predicting a line on the parametric $\left(m_V(0), m_A(0)\right)$ plane and a point on this line was chosen such that
it provides the least mean square deviation from the experimental data.

\begin{table}[!h]
\caption{\small The parameters of solution in the vector case.}
$\begin{array}{|c|c|c|c|c|c|c|c|c|c|}
 \hline
 \text{M, MeV} & A_F^V & A_F^A & B_F & A_m^V,\,\text{MeV}^2 & A_m^A,\,\text{MeV}^2 & B_m & L_{10}\text{, }10^{-3}& \text{$\Delta $m}_{\pi },\,\text{MeV} \\
 \hline
 925 & 0.001 & -0.003 & 0.743 & -(512)^2 & (765)^2 & 1.024 & -6.503 & 4.817 \\
 \hline
 1215 & 0.003 & 0. & 0.341 & -(940)^2 & -(191)^2 & 0.392 & -6.638 & 5.436 \\
 \hline
\end{array}$
\end{table}
\begin{table}[!h]
\caption{\small The mass spectrum of vector mesons (here and further in MeV).}
    $\begin{array}{|c|c|c|c|c|c|}
 \hline
 B_m & m_V\text{(0)} & m_V\text{(1)} & m_V\text{(2)} & m_V\text{(3)} & m_V\text{(4)}\\
 \hline
 1.024 & 770 & 1420 & 1825 & 2146 & 2422 \\
 \hline
 0.392 & 770 & 1461 & 1893 & 2229 & 2512 \\
 \hline
 \text{Exp.} & 775 & 1465\pm25 & 1909\pm17\pm25 & 2254\pm22 & \text{---} \\
 \hline
\end{array}$
\end{table}
\begin{table}[!h]
\caption{\small The mass spectrum of axial mesons.}
$\begin{array}{|c|c|c|c|c|c|}
 \hline
 B_m & m_A\text{(0)} & m_A\text{(1)} & m_A\text{(2)} & m_A\text{(3)} & m_A\text{(4)} \\
 \hline
 1.024 & 1200 & 1523 & 1855 & 2155 & 2425 \\
 \hline
 0.392 & 1200 & 1645 & 1992 & 2287 & 2547 \\
 \hline
 \text{Exp.} & 1230\pm40 & 1647\pm22 & 1930^{+30}_{-70} & 2270^{+55}_{-40} & \text{---} \\
 \hline
\end{array}$
\end{table}
\begin{table}[!h]
\caption{\small The predicted values of constants $F_{V,A}$.}
        $\begin{array}{|c|c|c|c|c|c|c|c|c|c|c|}
 \hline
 B_m & F_V\text{(0)} & F_V\text{(1)} & F_V\text{(2)} & F_V\text{(3)} & F_V\text{($\infty $)} & F_A\text{(0)} & F_A\text{(1)} & F_A\text{(2)} & F_A\text{(3)} & F_A\text{($\infty $)} \\
 \hline
 1.024 & 138 & 135 & 133 & 133 & 132 & 116 & 125 & 129 & 130 & 132 \\
 \hline
 0.392 & 144 & 141 & 138 & 136 & 132 & 133 & 133 & 133 & 132 & 132 \\
 \hline
\end{array}$
\end{table}

\begin{figure}
    \includegraphics[width=0.985\linewidth,fbox]{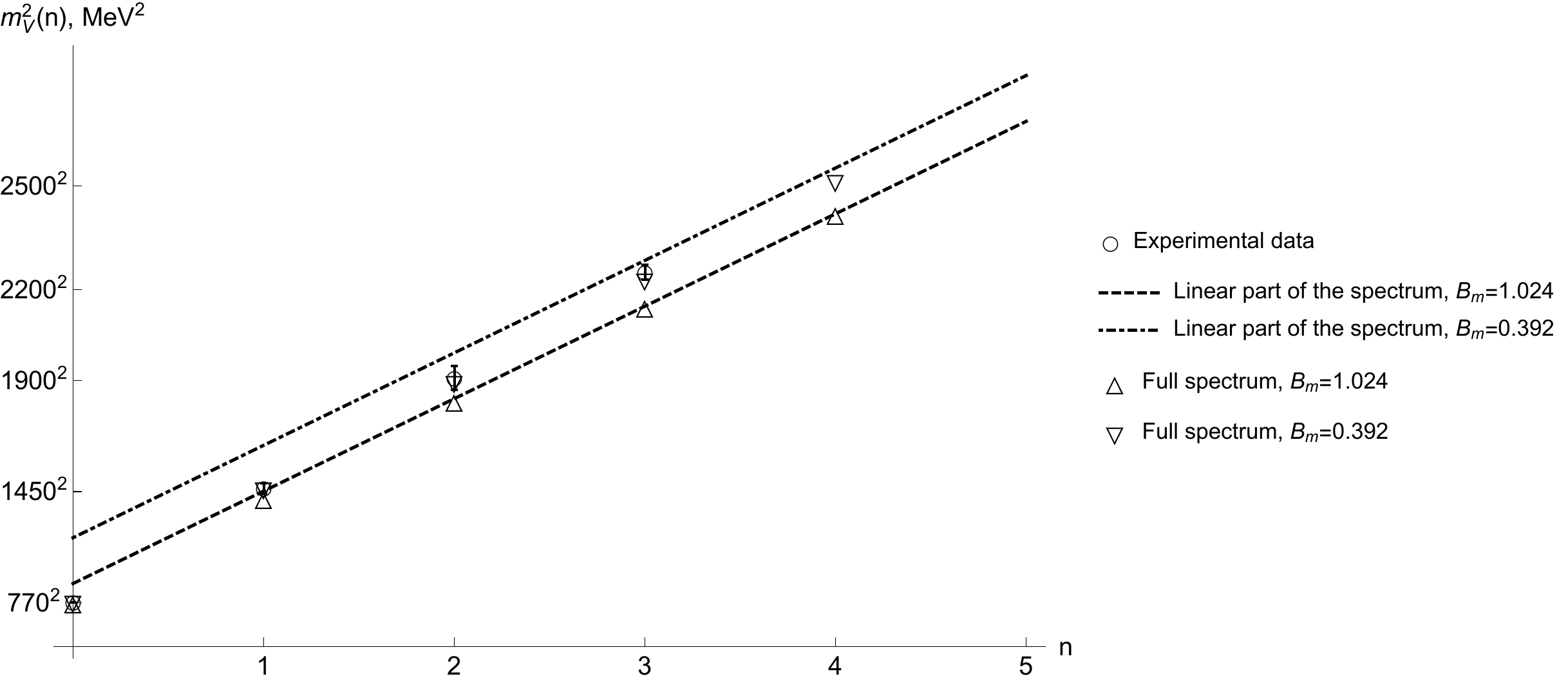}
    \caption{A graphical comparison of predicted and experimental vector spectrum.}
\end{figure}
\begin{figure}
    \includegraphics[width=0.985\linewidth,fbox]{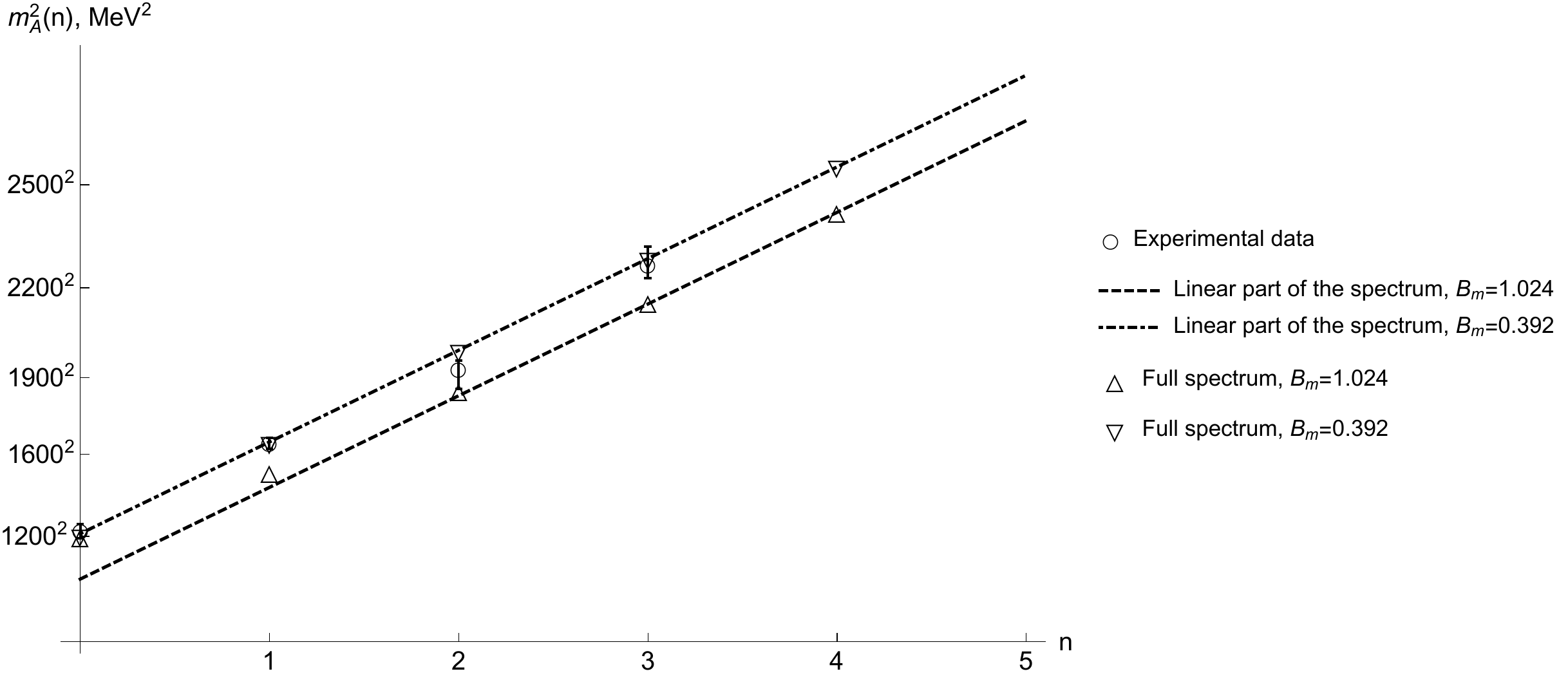}
    \caption{A graphical comparison of predicted and experimental axial spectrum.}
\end{figure}

\subsection{Scalar and pseudoscalar mesons}

The mass spectrum ansatz~\eqref{22}--\eqref{23} contains 9 parameters. We have 4 sum rules~\eqref{25}--\eqref{28}.
By assumption, the slope $a$ and exponent $B_m$ are universal parameters and we take them from the vector channel.
Thus we need 3 additional constraints. Following Ref.~\cite{we} these constraints will be the masses
of three states: $m_S(0)=1000\,\text{MeV}$, $m_P(0)=0$, $m_P(1)=1300\,\text{MeV}$.
Since the vector channel had two solutions with different values of exponent $B_m$,
we will have two solutions in the scalar channels corresponding to two different fixations of $B_m$.
In addition, we will consider separately the possibilities $\pi$-in and $\pi$-out.
The calculated spectrum for all four variants is displayed and compared with
the experimental data for $f_0$ and $\pi$ mesons in Tables~5-12 and Plots~3-6. When there
are two possible assignments for the predicted excited scalar mesons we show
two lines for the experimental data in Tables~6 and~10.

\begin{table}[h!]
    \caption{\small The parameters of solution in the scalar case $\pi$-in.}
$\begin{array}{|c|c|c|c|c|c|c|c|}
 \hline
 B_m & \bar{M},\,\text{MeV} & A_m^P,\,\text{MeV}^2 & A_m^S,\,\text{MeV}^2 & A_G^P & A_G^S & B_G & L_8\text{, }10^{-3} \\
 \hline
 1.024 & 824 & -(824)^2 & (566)^2 & 0.004 & -0.003 & 0.667 & 0.725 \\
 \hline
 0.392 & 1159 & -(1159)^2 & -(586)^2 & 0.006 & 0.002 & 0.308 & 0.761 \\
 \hline
\end{array}$
\end{table}
\begin{table}[h!]
    \caption{\small The mass spectrum of scalar mesons in the $\pi$-in case.}
    $\begin{array}{|c|c|c|c|c|c|}
 \hline
 B_m & m_S\text{(0)} & m_S\text{(1)} & m_S\text{(2)} & m_S\text{(3)} & m_S\text{(4)} \\
 \hline
 1.024 & 1000 & 1431 & 1797 & 2111 & 2388 \\
 \hline
 0.392 & 1000 & 1538 & 1922 & 2236 & 2508 \\
 \hline
 \multirow{2}{*}{\text{Exp.}} & \multirow{2}{*}{$990 \pm 20$} & 1200\text{--}1500 & 1723^{+6}_{-5} & 2101 \pm 7 & 2314 \pm 25 \\
 \cline{3-6}
 & & 1504 \pm 6 & 1992 \pm 16 & 2189 \pm 13 & 2539 \pm 14^{+38}_{-14} \\
 \hline
\end{array}$
\end{table}
\begin{table}[h!]
    \caption{\small The mass spectrum of pseudoscalar mesons in the $\pi$-in case.}
    $\begin{array}{|c|c|c|c|c|c|}
 \hline
 B_m & m_P\text{(0)} & m_P\text{(1)} & m_P\text{(2)} & m_P\text{(3)} & m_P\text{(4)} \\
 \hline
 1.024 & 0 & 1300 & 1761 & 2100 & 2385 \\
 \hline
 0.392 & 0 & 1300 & 1800 & 2166 & 2466 \\
 \hline
 \text{Exp.} & \text{---} & 1300\pm100 & 1812\pm12 & 2070\pm35 & 2360\pm25 \\
 \hline
\end{array}$
\end{table}
\begin{table}[h!]
    \caption{\small The predicted values of constants $G_{S,P}$ in the case $\pi$-in.}
    $\begin{array}{|c|c|c|c|c|c|c|c|c|c|c|}
 \hline
 B_m & G_P\text{(0)} & G_P\text{(1)} & G_P\text{(2)} & G_P\text{(3)} & G_P\text{($\infty $)} & G_S\text{(0)} & G_S\text{(1)} & G_S\text{(2)} & G_S\text{(3)} & G_S\text{($\infty $)} \\
 \hline
 1.024 & 192 & 186 & 183 & 181 & 179 & 169 & 174 & 177 & 178 & 179 \\
 \hline
 0.392 & 198 & 194 & 190 & 187 & 179 & 186 & 184 & 183 & 182 & 179 \\
 \hline
\end{array}$
\end{table}
\begin{figure}
    \includegraphics[width=0.985\linewidth,fbox]{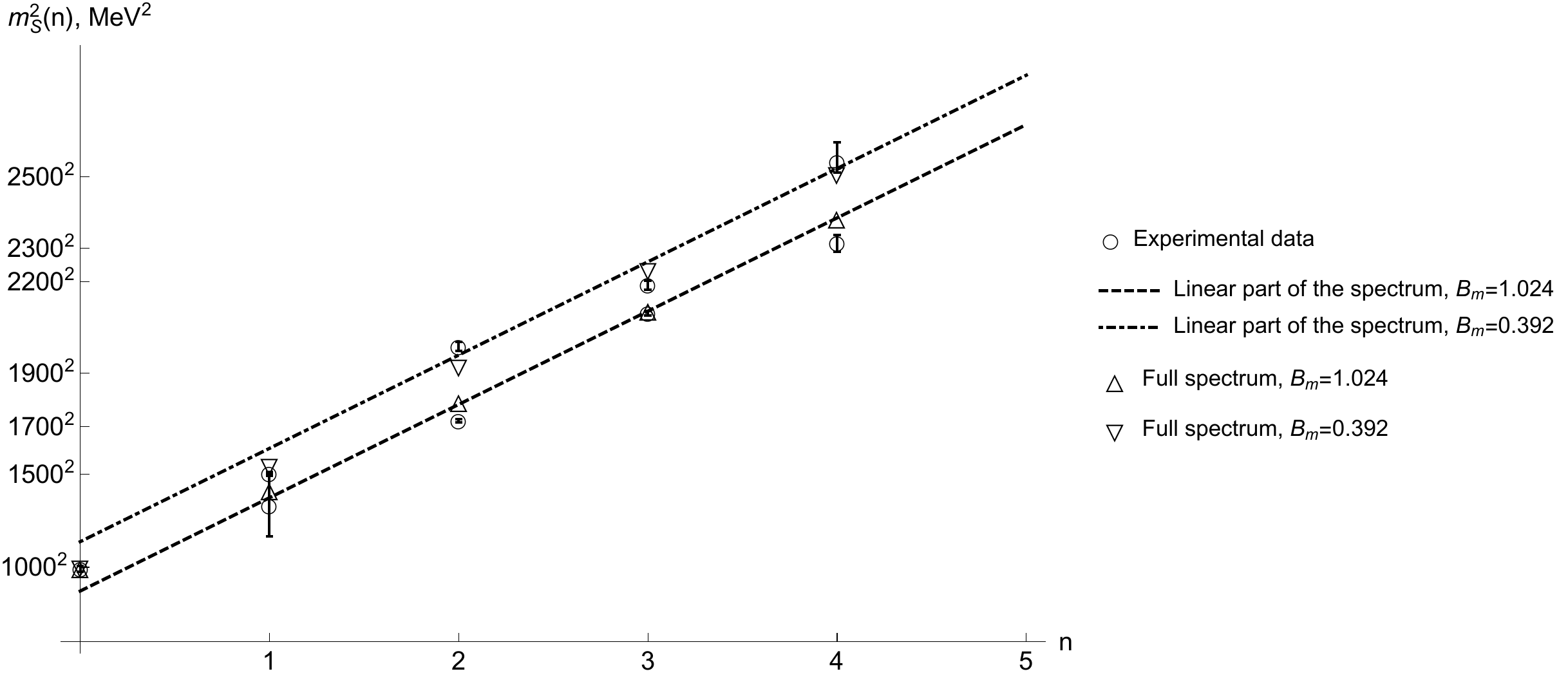}
    \caption{\small A graphical comparison of predicted and experimental scalar spectrum in the case $\pi$-in.}
\end{figure}
\begin{figure}
    \includegraphics[width=0.985\linewidth,fbox]{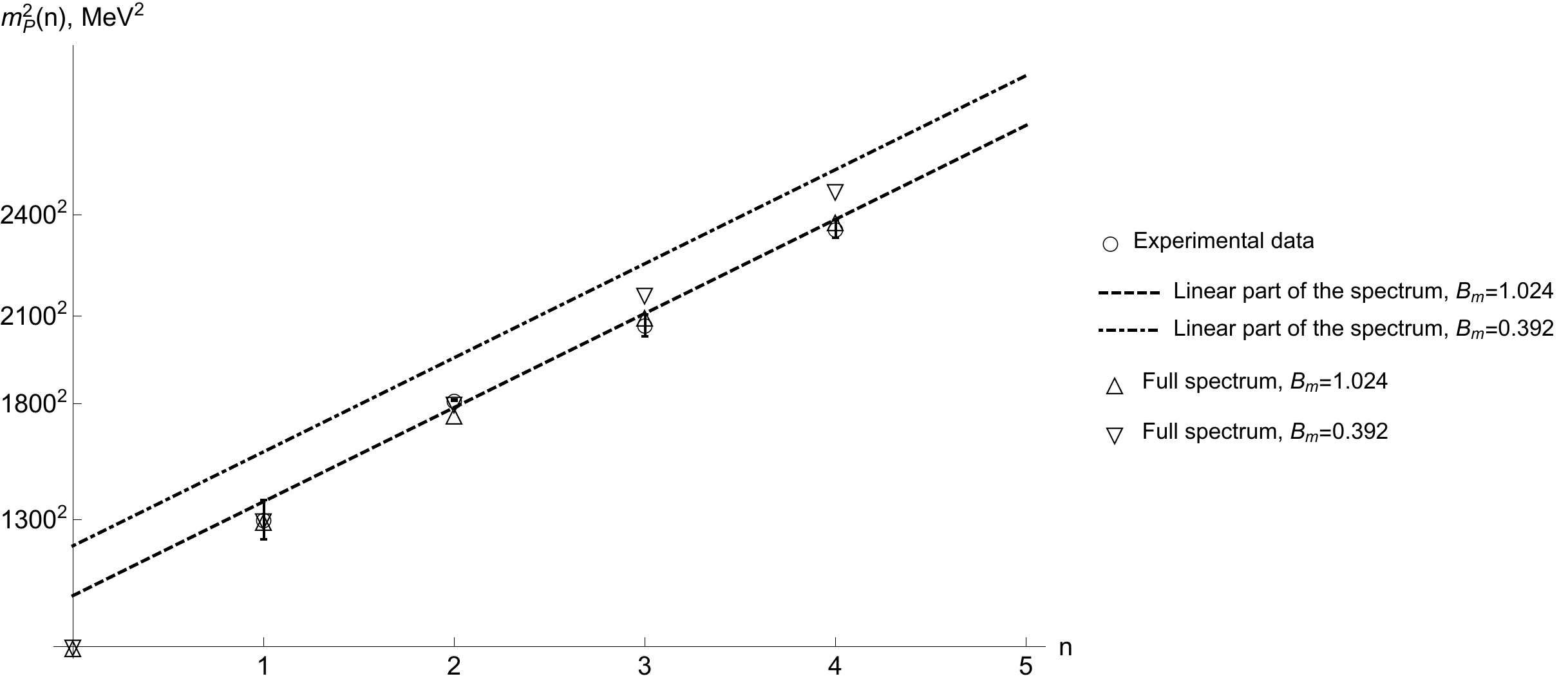}
    \caption{\small A graphical comparison of predicted and experimental pseudoscalar spectrum in the case $\pi$-in.}
\end{figure}

\begin{table}[!h]
    \caption{\small The parameters of solution in the scalar case $\pi$-out.}
    $\begin{array}{|c|c|c|c|c|c|c|c|}
 \hline
 B_m & \bar{M},\,\text{MeV} & A_m^P,\,\text{MeV}^2 & A_m^S,\,\text{MeV}^2 & A_G^P & A_G^S & B_G & L_8\text{, }10^{-3} \\
 \hline
 1.024 & 1467 & -(679)^2 & -(1073)^2 & 0.008 & 0.025 & 1.181 & 1.147 \\
 \hline
 0.392 & 1613 & -(955)^2 & -(1266)^2 & 0.007 & 0.012 & 0.408 & 0.747 \\
 \hline
\end{array}$
\end{table}
\begin{table}[!h]
    \caption{\small The mass spectrum of scalar mesons in the $\pi$-out case.}
    $\begin{array}{|c|c|c|c|c|c|}
 \hline
 B_m & m_S\text{(0)} & m_S\text{(1)} & m_S\text{(2)} & m_S\text{(3)} & m_S\text{(4)} \\
 \hline
 1.024 & 1000 & 1730 & 2124 & 2421 & 2674 \\
 \hline
 0.392 & 1000 & 1665 & 2093 & 2423 & 2699 \\
 \hline
 \multirow{2}{*}{\text{Exp.}} & \multirow{2}{*}{$990 \pm 20$} & \multirow{2}{*}{$1723^{+6}_{-5}$} & 2189 \pm 13 & \multirow{2}{*}{$2314\pm25$} & \multirow{2}{*}{---} \\
 \cline{4-4}
 &&& 2101 \pm 7 & & \\
 \hline
\end{array}$
\end{table}
\begin{table}[!h]
    \caption{\small The mass spectrum of pseudoscalar mesons in the $\pi$-out case.}
    $\begin{array}{|c|c|c|c|c|c|}
 \hline
 B_m & m_P\text{(0)} & m_P\text{(1)} & m_P\text{(2)} & m_P\text{(3)} & m_P\text{(4)} \\
 \hline
 1.024 & 1300 & 1800 & 2145 & 2428 & 2676 \\
 \hline
 0.392 & 1300 & 1800 & 2167 & 2466 & 2726 \\
 \hline
 \text{Exp.} & 1300\pm100 & 1812\pm12 & 2070\pm35 & 2360\pm25 & \text{---} \\
 \hline
\end{array}$
\end{table}
\begin{table}[!h]
    \caption{\small The predicted values of constants $G_{S,P}$ in the case $\pi$-out.}
    $\begin{array}{|c|c|c|c|c|c|c|c|c|c|c|}
 \hline
 B_m & G_P\text{(0)} & G_P\text{(1)} & G_P\text{(2)} & G_P\text{(3)} & G_P\text{($\infty $)} & G_S\text{(0)} & G_S\text{(1)} & G_S\text{(2)} & G_S\text{(3)} & G_S\text{($\infty $)} \\
 \hline
 1.024 & 205 & 188 & 182 & 180 & 179 & 252 & 205 & 187 & 182 & 179 \\
 \hline
 0.392 & 202 & 195 & 190 & 186 & 179 & 218 & 206 & 197 & 192 & 179 \\
 \hline
\end{array}$
\end{table}
\begin{figure}
    \includegraphics[width=0.985\linewidth,fbox]{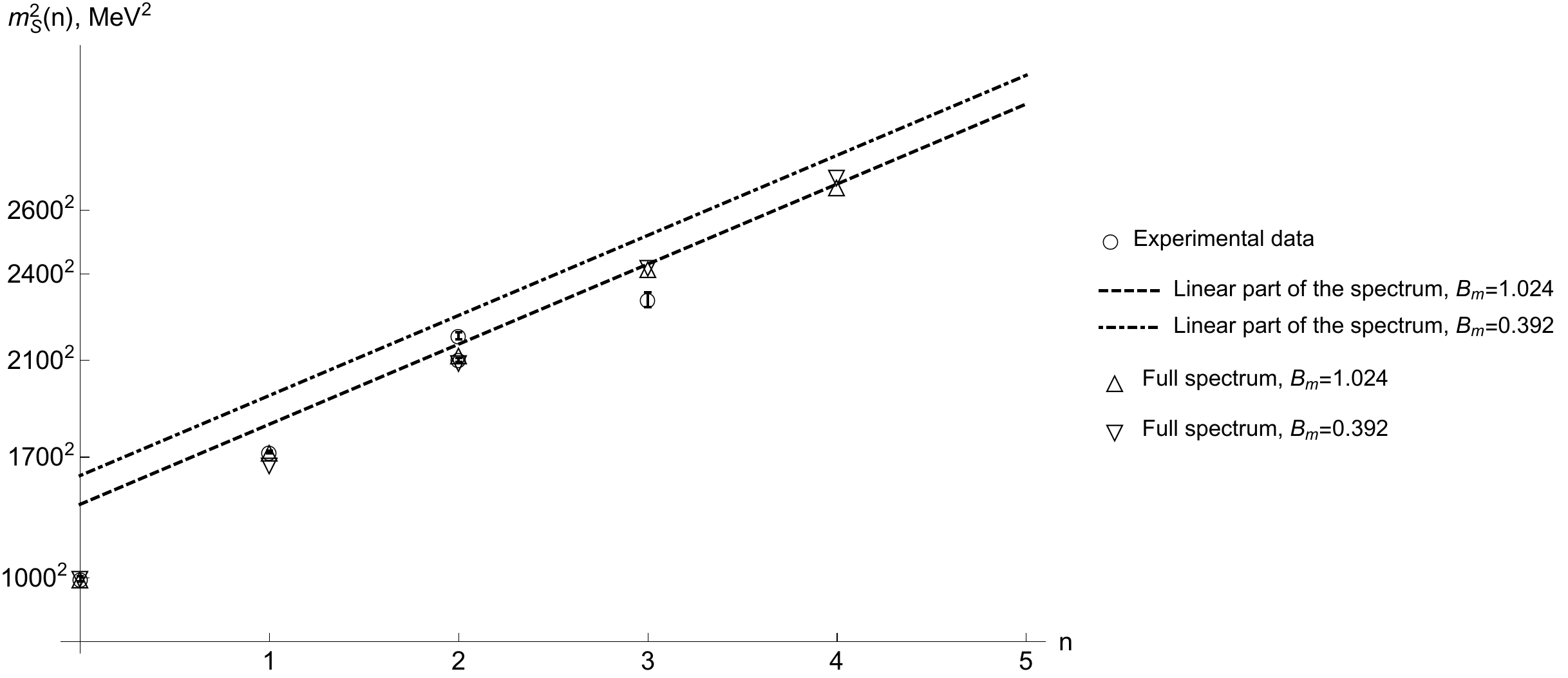}
    \caption{A graphical comparison of predicted and experimental scalar spectrum in the case $\pi$-out.}
\end{figure}
\begin{figure}
    \includegraphics[width=0.985\linewidth,fbox]{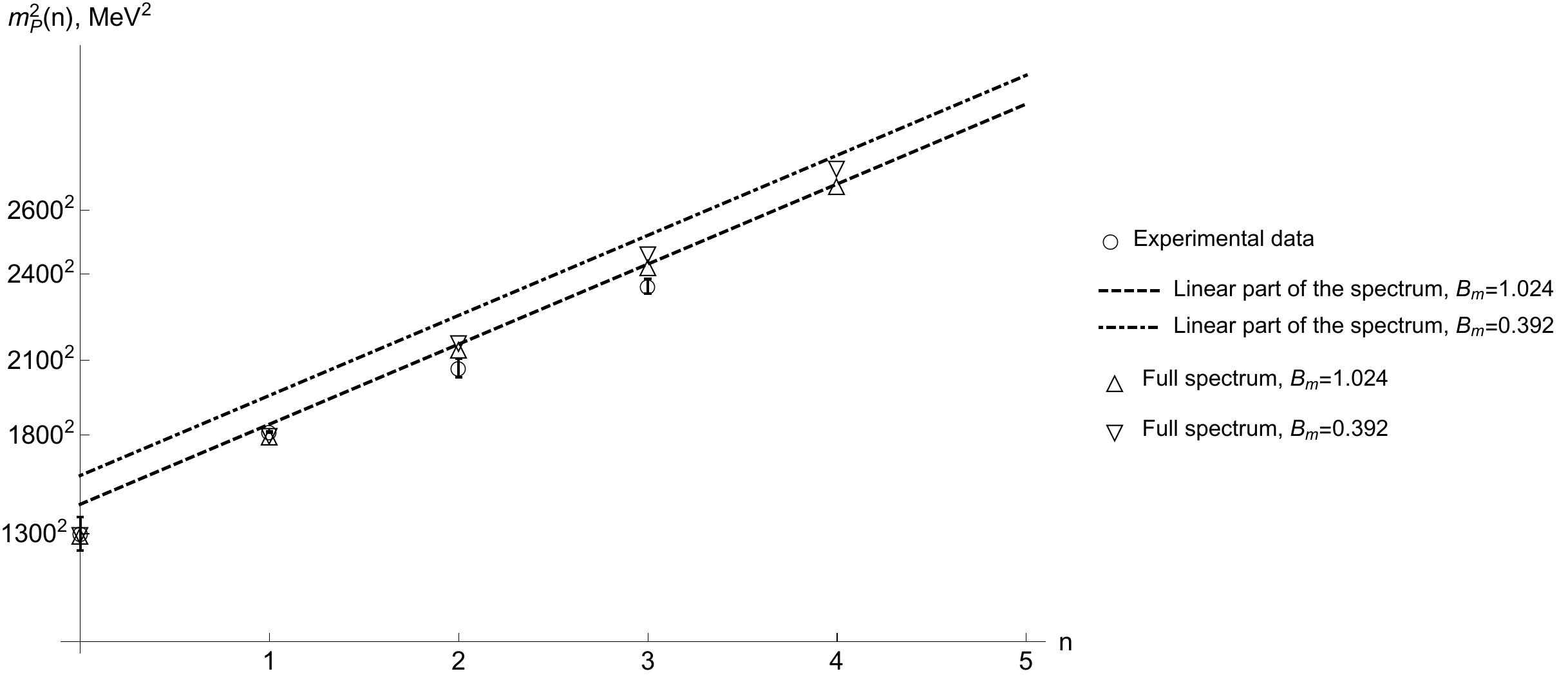}
    \caption{A graphical comparison of predicted and experimental pseudoscalar spectrum in the case $\pi$-out.}
\end{figure}

The numerical effect of terms in Eqs.~\eqref{25}--\eqref{28} displayed in bold face (missed in Ref.~\cite{we})
turned out to be small, far beyond the accuracy of the large-$N_c$ limit.

\section{Further remarks}

In the QCD sum rules, the quark and gluon condensates are external phenomenological parameters.
It is interesting to check the impact of their values on our numerical solutions. Such a check
was not done in Ref.~\cite{we}. We considered three radical possibilities: $\qqb=0$,
$\gmv=0$, and $\qqb=\gmv=0$. The caused shifts in numerical results turned out to be
at the level of 1\% or less, i.e. much less than the expected accuracy of our method based on the large-$N_c$ limit.

The question may appear why we used the input $f_{\pi}=103\,\mbox{MeV}$ which slightly differs from
the corresponding phenomenological value $f_{\pi}=93\,\mbox{MeV}$? The reason is that for lower values of $f_{\pi}$
our set of sum rules does not have any physical solutions. This feature was noticed in Ref.~\cite{we} and was confirmed
by our present simulations.

The scalar resonance $f_0(500)$ is known to be very controversial (see, e.g. "Note on scalar mesons below 2 GeV" in~\cite{pdg}
and recent review~\cite{pelaez}). It is interesting thus to check whether our model is able to describe this state.
We will proceed in the following way. The value of $m_S(0)$ will be decreased up to a point where the solutions cease to exist.
The scalar and pseudoscalar mass spectrum for this minimal value of $m_S(0)$ is presented in Table~13.
\begin{table}[!h]
    \caption{\small The scalar and pseudoscalar mass spectrum for minimal value of $m_S(0)$.}
    $
\begin{array}{|c|c|c|c|c|c|c|c|c|c|c|c|}
 \hline
 {} & B_m & m_S\text{(0)} & m_S\text{(1)} & m_S\text{(2)} & m_S\text{(3)} & m_S\text{(4)} & m_P\text{(0)} & m_P\text{(1)} & m_P\text{(2)} & m_P\text{(3)} & m_P\text{(4)} \\
 \hline
 \multirow{2}{*}{$\pi$-in} & 1.024 & 800 & 1386 & 1784 & 2107 & 2387 & 0 & 1300 & 1761 & 2100 & 2385 \\
 \cline{2-12}
 {}& 0.392 & 700 & 1422 & 1861 & 2201 & 2487 & 0 & 1300 & 1800 & 2166 & 2466 \\
 \hline
  \multirow{2}{*}{$\pi$-out} & 1.024 & 1000 & 1730 & 2124 & 2421 & 2674 & 1300 & 1800 & 2145 & 2428 & 2676 \\
 \cline{2-12}
 {}& 0.392 & 800 & 1591 & 2053 & 2400 & 2685 & 1300 & 1800 & 2167 & 2466 & 2726 \\
 \hline
\end{array}
$
\end{table}
It is seen that the second solution permits a lower value of $m_S(0)$. In any case there is no solution near 500~MeV.

Our ansatz for non-linear corrections distorts the linear trajectories in one direction.
This may look unnatural because one might expect that the physical masses are scattered uniformly near both sides of
straight radial Regge trajectory. The given feature may be modeled by adding extra factor $(-1)^n$ in front of the exponential
correction which makes this correction sign-alternating,
\begin{equation}
\label{40}
m^2_{V,A}(n)=M^2+a n+(-1)^n A_m^{V,A} e^{-B_m n}.
\end{equation}
This ansatz leads to the following sum rules in the vector sector (the changes caused by
the sign-alternating modification of ansatz are shown in bold face):

\noindent The sum rule at $1/Q^2$,
$$
\frac{1}{8 \pi^2}\lbbr 1+\frac{\alpha_s}{\pi}\rbbr =C.
$$
\noindent The sum rules at $1/Q^4$,
\begin{align}
a\lbbr C+A_F^V\rbbr -C\lbbr \frac{a}{2}+M^2\rbbr +A_F^V\Delta_F^{(1)}&=0,\\
a\lbbr C+A_F^A\rbbr -C\lbbr \frac{a}{2}+M^2\rbbr +A_F^A\Delta_F^{(1)}&=-f_\pi^2.
\end{align}
The sum rules at $1/Q^6$,
\begin{multline}
-2a\lbbr C+A_F^V\rbbr \lbbr M^2+A_m^V\rbbr +C\lbbr M^4+M^2 a+\frac{a^2}{6}\rbbr + \mathbf{\frac{2 C A_m^V}{e^{B_m}+1}}\\ - 2 A_F^V \Delta_F^{(2)}=\frac{\alpha_s}{12\pi}\gmv,
\end{multline}
\begin{multline}
-2a\lbbr C+A_F^A\rbbr \lbbr M^2+A_m^A\rbbr +C\lbbr M^4+M^2 a+\frac{a^2}{6}\rbbr + \mathbf{\frac{2 C A_m^A}{e^{B_m}+1}}\\ - 2 A_F^A \Delta_F^{(2)}=\frac{\alpha_s}{12\pi}\gmv.
\end{multline}
The sum rule at $1/Q^8$ in $\Pi^V(Q^2)-\Pi^A(Q^2)$,
\begin{multline}
3a\lbbr C+A_F^V\rbbr \lbbr M^2+A_m^V\rbbr ^2-3a\lbbr C+A_F^A\rbbr \lbbr M^2+A_m^A\rbbr ^2+\\
\mathbf{6 C \lbbr A_m^A-A_m^V\rbbr\lbbr\frac{M^2}{e^{B_m}+1}+\frac{a e^{B_m}}{\lbbr e^{B_m} + 1\rbbr^2}\rbbr}\\ +3(A_F^V-A_F^A)\Delta_F^{(3)}=-12\pi \alpha_s \qqb^2.
\end{multline}
Our numerical simulations, however, did not result in any physically reasonable solution.

\section{Conclusions}

The spectroscopic models based on the planar QCD sum rules remain
a viable non-perturbative approach to the spectroscopy of light mesons.

In the given work, we critically reassessed the analysis of radial Regge
trajectories with non-linear corrections performed in Ref.~\cite{we}
for the vector, axial, scalar, and pseudoscalar trajectories. Some errors
in the expressions for the sum rules in the scalar channel and for the
electromagnetic pion mass difference were found. Their influence on the
numerical solutions, however, is negligible. The main finding was the
discovery of the second solution in the considered planar sum rules.
The global description of the experimental data for vector and axial
mesons happens to be better in the case of the second solution.
In the scalar and pseudoscalar sectors, the quality of description
seems to be comparable for both solutions.

We confirmed the conclusion of Ref.~\cite{we} that the lowest scalar meson
cannot be made significantly lighter 1~GeV within the considered approach.
We checked the dependence of numerical results on the values of vacuum
condensates in the OPE and found it negligible within the accuracy of
the given large-$N_c$ method. The planar sum rules for a sign-alternating
exponential correction to the radial trajectories were derived and analyzed
numerically. We could not detect any physically acceptable numerical solution
for this ansatz. We observed an extremely strong sensitivity of electromagnetic
pion mass difference to the masses of lowest vector and axial mesons.
The fact that the physical value of this difference is achieved with
input masses very close to their central experimental values seems to
represent a nontrivial and interesting result.

It could be interesting to extend the present analysis to sectors with
other quantum numbers and to light mesons with open and hidden strangeness.

\end{document}